\documentclass[10pt]{article}
\usepackage[utf8]{inputenc} 
\usepackage[textwidth = 530 pt, textheight = 650 pt]{geometry}
\usepackage[table]{xcolor}
\usepackage{hyperref,amsthm,amsmath,amssymb,slashed,graphicx,bbold,fancybox,mathtools,color,tikz-cd,caption,subcaption,geometry,marginnote,xcolor,fancybox,verbatim, marginnote}
\usepackage{mathrsfs}
\usepackage{simplewick}
\usepackage{cite, mcite}
\usepackage{microtype}

\usepackage{tikz-cd,tikz}
\usetikzlibrary{arrows,decorations.markings} 

\usepackage[tikz]{mdframed}

\newcommand{\Comment}[1]{{}}

\Comment{\usepackage{xcolor}
\definecolor{MyDarkBlue}{rgb}{0.15,0.15,0.45}
\usepackage[linktocpage=true]{hyperref}
\usepackage{makeidx}
\usepackage{hypernat}}
\usepackage{framed}
\usetikzlibrary{mindmap,backgrounds}

\definecolor{MyDarkBlue}{rgb}{0.15,0.15,0.45}
\definecolor{shadecolor}{rgb}{0.85,0.85,0.85}
\definecolor{link-blue}{rgb}{0.15,0.15,0.65}
\definecolor{link-red}{rgb}{0.8,0.15,0.1}
\definecolor{link-green}{rgb}{0.15,0.50,0.15}
\definecolor{link}{rgb}{0.45,0.18,0.22}

\usepackage{graphicx}

\def\d{\partial}

\def\bpm{\begin{pmatrix}}
\def\epm{\end{pmatrix}}

\newcommand{\be}{\begin{equation}}
\newcommand{\ee}{\end{equation}}
\newcommand{\bse}{\begin{subequations}}
\newcommand{\ese}{\end{subequations}}
\newcommand{\bea}{\begin{eqnarray}}
\newcommand{\eea}{\end{eqnarray}}
\newcommand{\dd}{\mathrm{d}}

\hypersetup{
pdftitle={Remarks on BMS(3) invariant field theories},
pdfsubject={High Energy Physics},
pdfauthor={Thiago Araujo},
pdfkeywords={gauge; susy; strings; fields; cft},
pdfsubject={hep-th},
colorlinks=true,linkcolor=link,citecolor=link,urlcolor=link,linktocpage
}

\begin{document}

\renewcommand{\thefootnote}{\fnsymbol{footnote}}

\makeatletter
\@addtoreset{equation}{section}
\makeatother
\renewcommand{\theequation}{\thesection.\arabic{equation}}

\rightline{}
\rightline{}


\begin{flushright}
APCTP Pre2018-002\\
\today
\end{flushright}   

\vspace{0.3cm}


\begin{center}
{\LARGE \bf{Remarks on BMS\(_3\) invariant field theories: correlation functions and nonunitary CFTs}}
\end{center}
 \vspace{1truecm}
\thispagestyle{empty} \centerline{
 {\large \bf { Thiago Araujo}}\;
}

\vspace{.3cm}
\centerline{{\it
Asia Pacific Center for Theoretical Physics,}}
\centerline{{\it Postech, Pohang 37673, Korea}}\vspace{.2cm}
\centerline{{\tt thiago.araujo@apctp.org}}

\vspace{0.8truecm}

\thispagestyle{empty}

\centerline{\bf Abstract}

\vspace{.4truecm}

\begin{center}
\begin{minipage}[c]{380pt}
{\noindent  
We use the isomorphism between the BMS${}_3$ and the $W(2,2)$ algebras to reconsider some generic aspects of CFTs with the BMS${}_3$ algebra defined as a chiral symmetry. For unitarity theories, it is known that the extended symmetry generator acts trivially, and the resulting theory is equivalent to a CFT with a Virasoro symmetry only. For nonunitary CFTs, we define an operator depending on a nilpotent variable, and we organize the Verma module through the action of this new operator. Finally, we find the conditions imposed by the modified Ward identity. 
}
\end{minipage}
\end{center}

\vspace{0.1cm}

\setcounter{page}{1}
\setcounter{tocdepth}{2}

\renewcommand{\thefootnote}{\arabic{footnote}}
\setcounter{footnote}{0}

\linespread{1.1}
\parskip 4pt


\section{Introduction}

In a harbinger of the AdS/CFT correspondence, Brown and Henneaux \cite{Brown:1986nw} showed that the naive asymptotic symmetry $SL(2, \mathbb{R})\times SL(2, \mathbb{R})$ of the AdS${}_3$ space is enhanced to two copies of the Virasoro algebra $\mathtt{Vir}\oplus \overline{\mathtt{Vir}}$. These insights gave the necessary framework for the modern developments in the AdS${}_3$/CFT${}_2$ duality, e.g. \cite{Kraus:2006wn, Witten:2007kt, Sfondrini:2014via,  Gaberdiel:2015wpo, Afshar:2017okz, Strominger:2017zoo}. Recently, after the observations relating the memory effect, soft theorems and asymptotic symmetries \cite{Strominger:2017zoo, Strominger:2013jfa}, an increasing interest in asymptotic flat spaces has emerged. 

The striking aspect of asymptotic flat spaces is the enhancement of their symmetry group to the Bondi-Metzner-Sachs (BMS) group \cite{Bondi:1962px, Sachs:1962zza, Barnich:2009se, Barnich:2011mi, Oblak:2016eij}. In three dimensions, the BMS${}_3$ algebra can be obtained from two copies of the Virasoro algebra, with generators $({\cal L}, \bar{\cal L})$ and central charges $(c, \bar{c})$ respectively, when we define the operators
\be 
L_m:= {\cal L}_m - \bar{\cal L}_{-m}\; , \quad M_m:= \frac{1}{r_0}({\cal L}_m - \bar{\cal L}_{-m})
\ee
and we assume that the AdS${}_3$ radius $r_0$ goes to infinity. The BMS${}_3$ algebra is given by
\be 
\label{bms3}
\begin{split} 
[ L_m, L_n] &= (m-n) L_{m+n} + \delta_{m+n,0}\frac{m(m^2-1)}{12}c_1\\
[ L_m, M_n] &= (m-n) M_{m+n} + \delta_{m+n,0}\frac{m(m^2-1)}{12}c_M\\
[ M_m, M_n] &= 0\; 
\end{split}
\ee
where $c_1= c- \bar{c}$ and $c_M=\frac{1}{r_0}c_L$. 

BMS algebra appears in several physical contexts, integrable models for example \cite{Fuentealba:2017omf, Barnich:2012rz}, but we are particularly interested in the study of some field theories that are invariant under this symmetry algebra. Theories of this type may arise through the so-called \emph{flat space holography} \cite{Fareghbal:2013ifa, Fareghbal:2014oba, Fareghbal:2014qga, Fareghbal:2015bxd, Bagchi:2016bcd, Bagchi:2016geg, Bagchi:2017cpu}. Evidently, one would like to extend to asymptotic flat spaces the AdS/CFT duality whose raison d'\^etre is to explain some remarkable connections between the symmetries of the AdS${}_{D+1}$ string theory solutions and $D$-dimensional field theories living in their conformal boundary \cite{Maldacena:1997re, Witten:1998qj}. 

As with the Virasoro case, the BMS${}_3$ symmetry is infinite dimensional and this fact is responsible for some surprising algebraic features. First of all, it is isomorphic to the two-dimensioanl (2D) \emph{Galilean-conformal algebra} (GCA), which is an In\"on\"u-Wigner contraction of the Virasoro algebra, see \cite{Bagchi:2009my, Bagchi:2009pe, Bagchi:2009ca, Bagchi:2010eg, Bagchi:2012cy, Campoleoni:2016vsh, Rasmussen:2017eus}. Therefore, we may regard these two algebras as the ``same,'' but the drawback of these constructions is that we do not have the holomorphicity properties usually available in meromorphic conformal field theories (CFTs) as defined initially by Belavin, Polyakov and Zamolodchikov \cite{Belavin:1984vu}. 

We try to remedy this situation using a third realization of the BMS${}_3$ algebra: it has been shown that it is also isomorphic to the $W(2,2)$ algebra \cite{Aizawa:2011ga, Rasmussen:2017eus, Adamovic2017, zhang2007, 2015Jiang, 2016Adamovic}, and that is the viewpoint we adopt in the present work. Roughly speaking, the $W(2,2)$ algebra is a ${\cal W}$ algebra defined by the stress-energy tensor and an additional spin-2 field $W(z)$. Similar ideas have been proposed in \cite{Kausch:1990vg}, but with the strong assumptions of that work, the additional spin-2 field was restricted to be the stress-energy tensor. We apply the isomorphism BMS${}_3\simeq W(2,2)$ to study $2$D CFTs with this algebra as a chiral symmetry, and then we try to use the powerful holomorphicity properties of meromorphic CFTs.

For example, in a generic $2$D CFT, we can solve the theory without the use of an action principle or even the equations of motion in a miraculous way known as the \emph{bootstrap approach}. This formalism is an outcome of what a quantum field theory is: the result of two pieces of data, \emph{correlation functions} and \emph{symmetries}. Correlation functions are generally defined from the dynamical details of a particular physical system, while the symmetries impose nontrivial relations among the correlation functions. Evidently, the main role in the bootstrap formalism is played by the infinite-dimensional algebra, which manifests itself as the Ward identities \cite{Belavin:1984vu, Moore:1988qv, Ginsparg:1988ui, Ketov:1995yd, DiFrancesco:1997nk, Gaberdiel:1999mc,  Schottenloher:2008zz, Blumenhagen:2009zz, Recknagel:2013uja, Ribault:2014hia}.

Before going any further, we should remark that although the flat space holography is, probably, a rich source of BMS${}_3$ invariant field theories, the CFTs we consider in this work are not expected to be dual to the asymptotically flat spaces that we have mentioned before. The main motivation in this text is toward the construction of CFTs with extended symmetries defined by another spin-2 field of which the modes satisfy the BMS${}_3$ algebra. Under this perspective, we should consider that these (elusive) theories are similar to $\mathfrak{sl}(n)$ Toda CFTs which are endowed with extended $\mathcal{W}_\ell$ algebra $\ell\geq 3$ as chiral symmetries; see for example \cite{Fateev:2007ab, Fateev:2008bm} and references therein.

The paper is organized as follows. In section 2, we review the isomorphism BMS${}_3\simeq W(2,2)$. Despite the fact that these results are not new, we try to write this discussion in a chiral CFT language. Additionally, we slightly change the hypothesis of \cite{Rasmussen:2017eus} and we do not assume that the zero-mode Lie algebra is necessarily semisimple \cite{Figueroa-OFarrill:1994liu}. We show that outside the semisimple Lie algebras realm it is possible to generate the centrally extended algebra through a Sugawara construction. In the same section, we briefly review the detailed analysis of null states in BMS${}_3$ field theories performed in \cite{Bagchi:2009pe}. We pay special attention to some kind of monster in the theory, which is necessary to the main proposal of this work in section 5. Namely, there are zero-norm states that are not null in the usual sense, since they are not, in principle, annihilated by all positive $W(2,2)$-algebra modes. 

In section 3, we study the Ward identities, and we will see that the constraints naively imposed by these identities to the correlation function, also impose that the central charge $c_M$ vanishes.  As we will argue along this work, a null central charge $c_M$ is the worst scenario that we can find; the resulting theory is equivalent to a CFT without extended chiral symmetry at all, and in this case the action of the generators $\{ W_n| n \in \mathbb{Z}\} $ is completely trivial. In section 4, we avoid the case $c_M=0$ in our theory by changing some of our hypotheses: we embrace the existence of the seminull states, and we accept the fate of nonunitarity.  We will see how we can organize the descendant fields into multiplets, and we also propose a modification of the Ward identity that takes into account the Jordan structure of the theory. We conclude in section 5 and discuss further research directions.

\section{BMS(3) as a chiral W(2,2) algebra}

In this section, we review the isomorphism between the BMS${}_3$ and the W(2,2) algebras \cite{zhang2007, 2015Jiang, 2016Adamovic, Banerjee:2015kcx}. We mainly follow the notation of \cite{Gaberdiel:1999mc}, but \cite{Belavin:1984vu, Moore:1988qv, Ginsparg:1988ui, Ketov:1995yd, DiFrancesco:1997nk, Gaberdiel:1999mc,  Schottenloher:2008zz, Blumenhagen:2009zz, Recknagel:2013uja, Ribault:2014hia} may be useful. Let us assume that we have a CFT living in the Riemann sphere $\mathbb{CP}^1$. Given two states $|v\rangle$ and $|w\rangle$ with conformal weights $h_v$ and $h_w$, respectively, the operator product expansion (OPE) between their corresponding vertex operators is
\be 
\label{ope}
V_v(z) V_w(\zeta) = V(V(|v\rangle; z-\zeta)|w\rangle; \zeta) = \sum_{n\leq h_w} V(V_n^v |w\rangle; \zeta)(z-\zeta)^{-n-h_v}\; .
\ee
Observe that the sum has an upper bound, and it arises from the fact that the theory cannot have any state with negative conformal dimension in a unitary theory. The OPE between a quasiprimary field  $W(z)$ of conformal dimension 2 and the stress-energy tensor reads
\be 
\begin{split}
\label{opetw}
T(z) W(\zeta) \sim & V(L_2|\phi_{2}\rangle; \zeta)(z-\zeta)^{-4} + V(L_1|\phi_{2}\rangle; \zeta)(z-\zeta)^{-3} +  V(L_0|\phi_{2}\rangle; \zeta)(z-\zeta)^{-2}\\
& +  V(L_{-1}|\phi_{2}\rangle; \zeta)(z-\zeta)^{-1}\; ,
\end{split}
\ee
where $|\phi_2\rangle$ is the state associated to the vertex operator $W(z)$. The commutator of their components is
\be 
\begin{split}
[L_m, W_n] = & \sum_{k=-1}^2 \genfrac{(}{)}{0pt}{0}{m+1}{m-k} V_{m+n}(L_k| \phi_2 \rangle)\\
 = & V_{n+m}(L_{-1}|\phi_2 \rangle) + (m+1)V_{n+m}(L_{0}|\phi_2 \rangle) + \frac{m(m+1)}{2}V_{n+m}(L_{1}|\phi_2 \rangle)\\
&  + \frac{m(m^2-1)}{6}V_{n+m}(L_{2}|\phi_2 \rangle)\; .
\end{split}
\ee
Given that $W(z)$ is quasiprimary, we have some simplifications:
\begin{itemize}
	\item[{\bf \S 1)}] The state $|\phi_2\rangle$ has conformal weight $2$. Then, $L_0|\phi_2\rangle= 2 |\phi_2\rangle$ and $L_1|\phi_2\rangle = 0$.
	\item[{\bf \S 2)}] It is easy to show that 
	\be 
	\begin{split}
	V(L_{-1}|\phi_2\rangle;z) & = \sum_{n\in \mathbb{Z}} V_n^{L_{-1} |\phi_2\rangle} z^{-n-3} = \partial V(|\phi_2\rangle;z)=\sum_{n\in \mathbb{Z}} (-n-2)W_n z^{-n-3}\; ;
	\end{split}
	\ee
	therefore,
	\be 
	V_n({L_{-1} |\phi_2\rangle})= -(n+2)W_n\; .
	\ee
	\item[{\bf \S 3)}] Finally, the state $L_2|\phi_2\rangle$ has conformal weight $h=0$. In fact,
	\be 
	L_0 (L_2 |\phi_2\rangle) = ([L_0, L_2] + L_2 L_0) |\phi_2\rangle \; .
	\ee
	Using now that $[L_0, L_2]=-2 L_2$ and $L_0|\phi_2\rangle =2 |\phi_2\rangle $, we have $L_0 (L_2 |\phi_2\rangle)=0$.  All in all, we conclude that $L_2 |\phi_2\rangle$ is proportional to the vacuum $|0\rangle$, and it is usually taken to be zero. In the present case, we write
	\be 
	L_2 |\phi_2\rangle := \frac{c_2}{2}|0\rangle\; .
	\ee
	Additionally, we know that $V_n(|0\rangle)=\delta_{n,0}$ .
\end{itemize}
Putting all these facts together, we have 
\be 
\label{w2-comm}
[L_m, W_n] = (m-n) W_{m+n} + \frac{c_2}{12}m(m^2-1)\delta_{n+m,0}\; .
\ee

Finally, it is clear from (\ref{ope}) that the general form of the OPE $W(z)W(\zeta)$ is still undetermined, but we can make a consistent choice
\be 
W(z)W(\zeta)  = 0\quad \Leftrightarrow\quad [W_m, W_n]=0\; .
\ee 
Relaxing this condition may be interesting from the physical viewpoint, but we keep this choice along this text.

\begin{shaded}{\bf Definition:} \emph{The $W(2,2)$ algebra is defined by the generators $\{L_m, W_n\ | \ m,n \in \mathbb{Z} \}$ satisfying the following commutation relations}
\begin{equation}
\label{w22-alg}
\begin{split}
&[L_m, L_n ] = (m-n)L_{m+n}+ \frac{c_1}{12}m(m^2-1)\delta_{m+n,0}\\
& [L_m, W_n ]  = (m-n)W_{m+n}+ \frac{c_2}{12}m(m^2-1)\delta_{m+n,0}\\
& [W_m, W_n]  =0 \; .
\end{split}
\end{equation}
\emph{The Verma module for the $W(2,2)$ algebra has been constructed in} \cite{zhang2007, 2015Jiang, 2016Adamovic} .
Furthermore, the algebras (\ref{bms3}) and (\ref{w22-alg}) are clearly isomorphic; therefore, we make the following identifications: $c_M\equiv c_2$ and $W(z)\equiv M(z)$.
\end{shaded}

The symmetry algebra above can be equivalently defined in terms of the holomorphic fields
\be 
T(z) = \sum_{n\in\mathbb{Z}} L_n z^{-n-2}  \quad \mathrm{and} \quad
W(z) = \sum_{n\in\mathbb{Z}} W_n z^{-n-2}
\ee
satisfying the OPEs,
\be 
\label{bms3-ope}
\begin{split}
& T(z) T(w)  \sim \frac{1}{2}\frac{c_1}{(z-w)^4} + \frac{2 T(w)}{(z-w)^2} + \frac{\d T(w)}{(z-w)}\\
& T(z) W(w) \sim \frac{1}{2}\frac{c_2}{(z-w)^4} + \frac{2 W(w)}{(z-w)^2} + \frac{\d W(w)}{(z-w)}\\
& W(z) W(w) \sim 0\; .
\end{split}
\ee
These expressions have been written down in \cite{Banerjee:2015kcx}, in which the authors studied a free-field realization of the BMS${}_3$ algebras in terms of a $\beta\gamma$-ghost system. In their construction, the $\beta\gamma$ system generates a BMS${}_3$ algebra with central charges $(c_1,c_2)=(26,0)$, and the value of $c_2$ may be changed by a twist. 

\subsection{Sugawara construction}

Regarding the BMS${}_3$ algebra as a Galilean-Virasoro algebra, the authors of \cite{Rasmussen:2017eus}
obtained the operators $W(z)$, with central charge $c_2=0$, from the same Lie algebra they built the energy-momentum tensor $T(z)$ through a Sugawara construction. Now, we would like to show that, even when we assume that the underlying Lie algebras that generate the operators $T(z)$ and $W(z)$ are different, and we shall see that depending on the properties of the Lie algebra, it is possible to generate a nontrivial central charge $c_2$.

Suppose that among the vertex operators there exists a complete set of currents
\be 
\mathfrak{g}=\{j^a(z)\equiv V(|v^a\rangle ; z)|L_0|v^a\rangle=|v^ a \rangle\; ,\ a=1, \cdots, q\}\; .
\ee 
Using (\ref{ope}), we write their OPE as
\be 
j^a(z) j^b(\zeta) \sim \frac{K Z^{ab}}{(z-\zeta)^2} + \frac{C^{ab}_{\phantom{ab}c}\ j^c(\zeta)} {(z-\zeta)} + (j^a j^b)(\zeta)\; ,
\ee
where $K$, $Z^{ab}$ and $C^{ab}_{\phantom{ab}c}$ are constants. The current algebra reads 
\be 
[j_m^a, j_n^b]= C^{ab}_{\phantom{ae} c} j^c_{m+n} + m K Z^{ab} \delta_{m+n,0}\; .
\ee 
We do not assume that the zero mode algebra $\mathfrak{g}_0$, $[j_0^a, j_0^b]= C^{ab}_{\phantom{ae} c} j^c_{0}$, is semisimple. Therefore, the structure constants $C^{ab}_{\phantom{ae} c}=-C^{ba}_{\phantom{ae} c}$ do not need to be completely antisymmetric, and we do not impose that $Z^{ab}$ is proportional to the Cartan-Killing form
\be 
\kappa^{ab}=\frac{1}{{\cal N}_0} C^{ae}_{\phantom{ae} c} C^{bc}_{\phantom{bc} e}\; ,
\ee 
where ${\cal N}_0$ is the Dynkin index. Finally, the nondegeneracy of the Cartan-Killing form \cite{fuchs} is not necessary in this setting. In other words, we allow the current algebra to be as arbitrary as possible.

We try to build the operator $W(z)$ from a Sugawara construction defined by the currents $\mathfrak{g}$ as
\be 
W(z)=\gamma X^{ab}(j^a j^b)(z)\; .
\ee
Evidently, $X^{ab}=X^{ba}$ and we allow the possibility $X^{ab} \propto \kappa^{ab}$, but we do not take it for granted.  

The properties of the matrices $X^{ab}$ and $Z^{ab}$ are determined from the OPEs 
\be 
W(z) W(\zeta) \quad \textrm{and}\quad T(z) W(\zeta)
\ee
and the constraints imposed by the BMS(3) algebra:
\begin{itemize}
		\item[\bf \S $1$) ] It is easy to see that the OPE (\ref{ope}) implies that the fields $j^a(z)$ are primaries. Then, their transformation is
		\bse
		\be 
		T(z) j^a(\zeta) \sim \frac{j^a(\zeta)}{(z-\zeta)^2} +  \frac{\partial j^a(\zeta)}{(z-\zeta)}\; .
		\ee
		\ese
		Now, we are able to calculate $T(z) W(\zeta)$ as
		\be 
		\begin{split}
		T(z)W(\zeta) 
		= & \frac{\gamma X^{ab}}{2 \pi i}
		\oint_{\zeta} \frac{d x}{x-\zeta} \left\{ \contraction{}{T(z)}{j^a(x)}{j^b(\zeta)}
		T(z) j^a(x) j^b(\zeta)+ \contraction{}{T(z)}{}{j^a(x)} T(z)j^a(x)j^b(\zeta) \right\}\\
		\sim &
		\frac{\gamma K X^{ab} Z^{ab}}{(z-\zeta)^4} + \frac{2 W(\zeta)}{(z-\zeta)^2} + \frac{\partial W(\zeta)}{(z-\zeta)}\; ,
		\end{split}
		\ee
		and we see that $c_2 = 2 \gamma K  X^{ab} Z^{ab}$.
		
	\item[\bf \S $2$) ] Now, we calculate the following OPE:
	\bse
	\be 
	W(z)j^c(\zeta) 
	=	 \frac{\gamma X^{ab}}{2 \pi i}
	\oint_{z} \frac{d x}{x-z} \left\{ \contraction{}{j^a(z)}{j^b(x)}{j^c(\zeta)}
	j^a(z) j^b(x) j^c(\zeta)+ \contraction{j^a(z) }{j^b(x)}{}{j^c(w)} j^a(z)j^b(x)j^c(\zeta) \right\}. 
	\ee
	Then,
	\be 
	\begin{split}
	W(z)j^c(\zeta) \sim & \frac{2 \gamma K X^{ab} Z^{ad} C^{bc}_{\phantom{ae} d} }{(z-\zeta)^3} + \frac{2\gamma \left( K Z^{ca}  X^{ae} + X^{ab} C^{bc}_{\phantom{ae} d} C^{ad}_{\phantom{ae} e} \right)  j^e(\zeta)  }{(z-\zeta)^2}  \\
	& + \frac{2\gamma  (K X^{ab} Z^{bc} \partial j^a(\zeta) +X^{ab} C^{bc}_{\phantom{ab}d} (j^a j^d)(\zeta) )}{(z-\zeta)}\; ,
	\end{split}
	\ee
	that is,
	\be
		W(z)j^c(\zeta)
	\equiv \frac{\lambda_{(1)}^c}{(z-\zeta)^3} + \frac{\lambda_{(2)}^{ca} j^a(\zeta) }{(z-\zeta)^2} + \frac{\lambda_{(2)}^{ca} \partial j^a(\zeta) +\lambda_{(3)}^{bac} (j^a j^c)(\zeta) }{(z-\zeta)}\; .
	\ee
	\ese
	From this expression it is easy to see that the last term is a bit problematic, since the OPE $W(z)W(\zeta)$ will give recursively cubic terms in the currents. For semisimple Lie algebras, the complete antisymmetry of the structure constant implies that this term vanishes trivially. In the present case, if we want to avoid the obvious antisymmetry choice, the most economical constraint is $\sum_b X^{ab} C^{bc}_{\phantom{ae} d}=0$. Therefore,
	\be 
	\lambda_{(1)}^b=0 \; , \qquad \lambda_{(3)}^{abc}=0\; .
	\ee
	
	\item[\bf \S $3$) ] Using these results, it is easy to compute the OPE:
	\bse
	\be 
	W(z)W(\zeta) 
	=   \frac{\gamma X^{cd}}{2 \pi i}
	\oint_{\zeta} \frac{d x}{x-\zeta} \left\{ \contraction{}{W(z)}{j^a(x)}{j^b(\zeta)}
	W(z) j^c(x) j^d(\zeta)+ \contraction{}{W(z) }{}{j^b(x)} W(z) j^c(x) j^d(\zeta) \right\}. 
	\ee
	Therefore, 
	\be 
	\begin{split} 
	 W(z)W(\zeta) \sim & \frac{\gamma K Z^{ab} X^{bc} \lambda_{(2)}^{ca}}{(z-\zeta)^4} + \frac{2 \gamma  X^{ac} \lambda_{(2)}^{cb} }{(z-\zeta)^2}(j^a j^b)(\zeta)  + \frac{\gamma X^{ac} \lambda_{(2)}^{cb}  }{z-\zeta} \partial (j^a j^b)(\zeta)\; ,
	\end{split}
	\ee
	\ese
	where the condition $\sum_b X^{ab}C^{bc}_{\phantom{ab}d}=0$ implies that  the term with $(z-\zeta)^{-3}$ vanishes trivially. The condition $W(z)W(\zeta)\sim 0$ is satisfied if $K=0$ regardless of the Lie algebra properties, and it gives vanishing central charge $c_2=0$, or if $\sum_{c,d} X^{ac} Z^{cd} X^{db}=0$, that gives nontrivial constraints to the Lie algebras. In the latter case, depending on the underlying details of the Lie algebra, we can generate a nonzero central charge \(c_2\).\footnote{I would like to thank the anonymous referee for pointing out a mistake in an earlier version of the paper.}
\end{itemize}

We have seen that the condition $W(z) W(\zeta)\sim 0$ was a bit arbitrary. We can change this OPE taking the condition $X=Z^{-1}$. Therefore,
\bse
\begin{align}
T(z) W(\zeta)& \sim \frac{\gamma K \dim(\mathfrak{g}) }{(z-\zeta)^4} + \frac{2 W(\zeta)}{(z-\zeta)^2} + \frac{\partial W(\zeta)}{z-\zeta}\\
W(z) W(\zeta) & \sim 2\gamma K\left( \frac{K \gamma \dim(\mathfrak{g})}{(z-\zeta)^4} + \frac{2 \gamma K}{(z-\zeta)^2} W(\zeta) + \frac{\gamma K}{z-\zeta} \partial W(\zeta) \right) = 2 \gamma K T(z) W(\zeta)\; .
\end{align}
\ese
We conclude that $W(z)=2 \gamma K T(z)$. In summary, given a stress-energy momentum tensor $T(z)$, possibly, built from a Sugawara approach, we can construct a second quasiprimary field $W(z)$ with vanishing central charge $c_2$ provided a nonsemisimple Lie algebra $\{j^a\}$ is given. 

\subsection{Null states}

The complete analysis of generic singular states in the BMS${}_3$ invariant field theories, has been performed in \cite{Bagchi:2009pe}, and the authors\footnote{Their analysis can be easily translated into our notation by the transformation $(\Delta,\xi, C_1, C_2)_{theirs}\equiv (h, h^{(2)}, c_1/12, c_2/12)_{ours}$.} were able to find null vector for special values of the CFT data $(h, h^{(2)}, c_1,c_2)$. In this subsection, we review their construction paying special attention to zero-norm states which appear in the representation theory of BMS${}_3$ invariant field theories.

We know that the highest-weight representation of a generic ${\cal W}_\ell$ algebra is defined by
\bse
\be 
\label{hws-01}
\begin{split}
& L_0|h,\ell\rangle = h |h,\ell\rangle \; , \quad W_0^{(\ell)}|h,h^{(\ell)}\rangle = \ell |h,h^{(\ell)}\rangle\\
& L_n|h,h^{(\ell)}\rangle = 0\; \; , \quad W_n^{(\ell)}|h,h^{(\ell)}\rangle = 0\; , \quad \forall \ n\geq 1\; ,
\end{split}
\ee
with Verma module
\be 
{\cal V}_{h, h^{(\ell)}}:= \mathrm{span} \left\{\left. \prod_{n_i \in \vec{n} } L_{-n_i} \prod_{ m_j\in \vec{m}} W_{-m_j} |h,h^{(\ell)}\rangle \right| n_i, m_j \in \mathbb{N}, n_{i+1}\geq n_i, m_{j+1}\geq m_j \right\}\; .
\ee 
As usual, there is a natural $L_0$ grading
\be 
{\cal V}^{(N)}_{h, h^{(\ell)}}:=\mathrm{span} \left\{\left. \prod_{n_i \in \vec{n} } L_{-n_i} \prod_{ m_j\in \vec{m}} W_{-m_j} |h,h^{(\ell)}\rangle  \in {\cal V}_{h, h^{(\ell)}} \right| \sum_i n_i + \sum_j m_j=N \right\}\; .
\ee
Then, the Verma module is written as
\be 
{\cal V}_{h, h^{(\ell)}}= \bigoplus_{ N\in \mathbb{Z}} {\cal V}_{h, h^{(\ell)}}^{(N)}\; .
\ee
\ese

In our case, the ${\cal W}$-algebra generator has $\ell=2$, and we write the highest-weight states as $|h\rangle\equiv |h, h^{(2)}\rangle$ with $L_0|h\rangle = h |h\rangle$ and $W_0|h\rangle = h^{(2)}|h\rangle$. Additionally, there may be states satisfying
\be 
L_0|\chi\rangle = (h + A) |\chi\rangle\; , \quad W_0|\chi\rangle = h^{(2)} |\chi\rangle\; , \quad L_n|\chi\rangle = 0 \; , \quad W_n|\chi\rangle = 0 \qquad \forall \; n\in \mathbb{N}^\ast\; ,
\ee
called \emph{null} or \emph{singular}, that are simultaneously primaries and descendants. Null states generate their own module  ${\cal V}_0$, also they are orthogonal to ${\cal V}_{h, h^{(2)}}$ and, in particular, to themselves, that is $\langle\chi| \chi \rangle=0$. The important point is that these vectors decouple from the set physical states, and the final space is a submodule generated by the quotient ${\cal V}_{h, h^{(2)}}/{\cal V}_0$ \cite{Ketov:1995yd, DiFrancesco:1997nk}. 

Using these results, we should observe that in the BMS${}_3$ algebra, any state of the form $|\chi\rangle=W_{-n}|h\rangle$ is a zero-norm state, that is,
\be 
||\chi||^2=\langle v| W_{n} W_{-n}|h \rangle = \langle v| W_{-n} W_{n}|v \rangle = 0\; .
\ee
On the other hand, these states are not, necessarily, annihilated by all the generators $L_m$, with $m\geq 1$. In other words, such states are not necessarily null. Acting with the Virasoro generators, we find
\bse
\be 
L_0 |\chi\rangle = L_0 (W_{-n}|h\rangle)=(n+h) |\chi\rangle\; ,
\ee
and
\be 
L_m  |\chi\rangle = L_m (W_{-n}|h\rangle) = 
\left\{
\begin{array}{ll}
m\left(2h^{(2)} +\frac{c_2}{12}(m^2-1)\right) |h\rangle &\; , \mathrm{if} \ \ m=n\\
0  &\; , \mathrm{if} \ \ m>n>0  \\
(m+n) W_{m-n}^{(2)} |h\rangle &\; , \mathrm{if} \ \ n>m>0  \\
\end{array}
\right.
\ee
\ese
Therefore, the condition
\be 
\label{null-cond}
2h^{(2)} +\frac{c_2}{12}(m^2-1) = 0
\ee
needs to be imposed. We need to be careful with the range $n>m>0$ since it is not zero, in principle. We say that zero-norm states that are not annihilated by all positive modes are \emph{seminull}. Therefore, we would like to see under which conditions the semi-null states are singular.

Moreover, it has been shown that if we assume (\ref{null-cond}) with $c_2\neq 0$, the null vector is unique, up to a multiplicative constant,  and it is of the form $W^{k_1}_{-n_1}\cdots W^{k_n}_{-n_n}|h\rangle$, and  that when $c_2=0$, we recover the Verma module of an ordinary CFT with a symmetry given by the Virasoro algebra only \cite{Bagchi:2009pe, 2015Jiang}. In this last case, the Verma module is reducible, and we can build usual null states using the generators $L_{-n}$. But observe that even in that case we still have the seminull states, and in this section we would like to pay attention to them.

	\paragraph{$\mathbf{N=1}:$} For the first level in the Verma module, we have two states 
	\be
	| \chi_1 \rangle = W_{-1}|h\rangle\quad \textrm{and}\quad | \chi_2 \rangle = L_{-1}|h\rangle.
	\ee 
	\begin{itemize}
	\item[{\it i})] The vector$|\chi_1\rangle = W_{-1}|h\rangle$ is a zero-norm state, that is $||\chi_1||^2=0$. Moreover, it is easy to see that 
	\be 
	L_0|\chi_1\rangle=(h+1) |\chi_1\rangle\; , \qquad W_0|\chi_1\rangle=h^{(2)}|\chi_1\rangle\quad \textrm{and}\quad 
	W_n|\chi_1\rangle = 0\quad  n\geq 1\; .
	\ee 
	In addition,
	\be 
	\begin{split}
	L_n |\chi_1\rangle=& \left((n+1)W_{n-1} +\frac{c_2}{12}n(n^2-1)\delta_{n-1,0} \right)|h\rangle \\
	=&  2 \delta_{n,1}h^{(2)}|h\rangle \; , \quad \forall\; n\in \mathbb{N}^\ast\; .
	\end{split}
	\ee
	We conclude that this state is seminull, unless we impose $h^{(2)}=0$. At this point, the condition (\ref{null-cond}) does not give $c_2=0$. Then, if $h^{(2)}=0$, $W_{-1}|v\rangle$ is  null.

	\item[{\it ii})] Consider now
	\be 
	|\chi_2\rangle = L_{-1}|h\rangle\; .
	\ee 
	As usual, we have $||\chi_2||^2=2 h$. In addition,
	\be 
	L_0 |\chi_2\rangle = (h+1)|\chi_2\rangle\; , \qquad W_0 |\chi_2\rangle = |\chi_1\rangle + h^{(2)}|\chi_2\rangle\; ,
	\ee
	and 
	\be 
	L_n |\chi_2\rangle = 2 h \delta_{n,1}|h\rangle\; , \qquad 	W_n |\chi_2\rangle = 2 h^{(2)} \delta_{n,1}|h\rangle\; .
	\ee
	The state is null just in the case $h=h^{(2)}=0$.
	\end{itemize}	
	
	These conditions can be derived from the $2\times 2$ Gram matrix\footnote{The order is twice the order of a theory with the Virasoro symmetry only.}
	\be
	G^{(1)} = 
	\begin{pmatrix}
	\langle h | L_1 L_{-1} |h \rangle & \langle h | W_1 L_{-1} |h \rangle\\
	\langle h | L_1 W_{-1} |h \rangle & \langle h | W_1 W_{-1} |h \rangle
	\end{pmatrix}	 
	\ee	
	whose determinant $\det G^{(2)}=-4 h^{(2)}$	vanishes just when $h^{(2)}=0$. Furthermore, this condition says that if $h^{(2)}>0$ we have a nonunitary theory.
	
\paragraph{$\mathbf{N=2}:$} At level 2, we can construct a generic vector of the form
\be 
\label{ba-null}
|\mu_1\rangle = \left( a_1 W_{-2} + a_2 L_{-2} + a_3 L_{-1} W_{-1} + a_4 W_{-1}^2 + a_5 L_{-1}^2  \right)|h\rangle\; ,	
\ee
and imposing the null vectors conditions, we determine constants $a_i$, see \cite{Bagchi:2009pe}. 

	\begin{itemize}
	\item[{\it i})] The textbook example of null vector at level $N=2$
\be 
|\mu_2\rangle = \left( L_{-2} - \frac{3}{2(2h+1)}L_{-1}^2 \right)|h\rangle\; ,
\ee
is seminull in our case, unless $h^{(2)}=0$.
	
	\item[{\it ii})] We also have a seminull state of the form
	\be 
	\label{null-st2}
	|\mu_3\rangle = \left(W_{-2}+ a_0 W_{-1} W_{-1}\right)|h\rangle\; ,
	\ee
	where $a_0$ is a constant to be determined. Obviously $||\mu_3 ||^2=0$ and $W_n|\mu_3\rangle=0$ . Additionally
	\be 
	L_0 |\mu_3\rangle = (h+2)|\mu_3\rangle \quad \mathrm{and} \quad W_0 |\mu_3\rangle = h^{(2)}_2|\mu_3\rangle\; .
	\ee
	Therefore, we need to calculate
	\be 
	\begin{split}
	L_n|\mu_3\rangle & = L_n (W_{-2}+ a_0 W_{-1} W_{-1})|h\rangle\\
	& = \left[ 
	(n+2)W_{n-2} + \frac{c_2}{12}n(n^2-1)\delta_{n-2,0}	\right. \\ 
	&\hspace{1.0cm} \left. + 2a_0\left( (n+1)W_{n-1}+ \frac{c_2}{12}n(n^2-1)\delta_{n-1,0} \right)  W_{-1}
	\right]|h\rangle\; .
	\end{split}
	\ee
	From this expression, we easily see that 
	\bse
	\be 
	L_m |h\rangle = 0 \; , \quad \forall\; n\in \mathbb{N}^\ast\; ,
	\ee
	is identically satisfied if $h^{(2)}= c_2=0$ or 
	\be 
	h^{(2)}_2 = -\frac{c_2}{8}\quad \mathrm{and}\quad \qquad a_0=\frac{6}{c_2}
	\ee
	\ese
	which is consistent with (\ref{null-cond}).
	\end{itemize}
	
	Again, these conditions can be derived from the $4 \times 4$ Gram matrix
	\bse
	\be
	G^{(2)} = 
	\begin{pmatrix}
	\langle h | L_2 L_{-2} |h \rangle & \langle h | L_2 L_{-1}^2 |h \rangle & \langle h | L_2 L_{-1} W_{-1} |h \rangle & \langle h | L_2 W_{-1}^2 |h \rangle\\
	\langle h | L_1^2 L_{-2} |h \rangle & \langle h | L_1^2 L_{-1}^2 |h \rangle & \langle h | L_1^2 L_{-1} W_{-1} |h \rangle & \langle h | L_{-1}^2 W_{-1}^2 |h \rangle\\
	\langle h | W_2 L_{-2} |h \rangle & \langle h | W_2 L_{-1}^2 |h \rangle & \langle h | W_2 L_{-1} W_{-1} |h \rangle & \langle h | W_2 W_{-1}^2 |h \rangle\\
	\langle h | W_{1}^2 L_{-2} |h \rangle & \langle h | W_{1}^2 L_{-1}^2 |h \rangle & \langle h | W_{1}^2 L_{-1} W_{-1} |h \rangle & \langle h | W_{1}^2 W_{-1}^2 |h \rangle\\	
	\end{pmatrix}	 
	\ee	
	so that
		\be
	G^{(2)} = 
	\begin{pmatrix}
	4h + c/2 &  6h & 6 h^{(2)} & 0\\
	6h & 4h(1 + 2h) & 4h^{(2)}(1 + 2h) & 8 (h^{(2)})^2\\
	4 h^{(2)}+c_2/2 & 6 h^{(2)} & 0 & 0\\
	0 & 8 (h^{(2)})^2 & 0 & 0 \\	
	\end{pmatrix}	 
	\ee	
	\ese
	whose determinant $\det G^{(2)}=192 (h^{(2)})^5(c_2 + 8 h^{(2)})$ vanishes when $h^{(2)}=0$ or $h^{(2)}=-c_2/8$.
	
	\paragraph{$\mathbf{N=3}:$} For completeness, at level 3 we have
	\be 
	|\nu \rangle = \left(W_{-3}+ a_1 W_{-2} W_{-1} + a_2 (W_{-1})^3\right)|h\rangle\; .
	\ee
	It is easy to see that $||\nu ||^2=0$,\ \ $W_n|\nu \rangle=0$ \ $\forall n>0$, and
	\be 
	L_0 |\nu\rangle = (h+3)|\nu\rangle\; , \qquad  	W_0 |\nu\rangle = h^{(2)} |\nu\rangle\; .
	\ee
	Furthermore,
	\be 
	\begin{split} 
	L_n|\nu\rangle  = & \left\{(n+3) W_{n-3} + \frac{c_2}{12}n(n^2-1)\delta_{n-3,0} \right.\\
	& +a_1 \left[(n+1)W_{n-1} +\frac{c_2}{12}n(n^2-1)\delta_{n-1,0}\right]  W_{-2} \\ 
	& + a_1 \left[(n+2)W_{n-2} +\frac{c_2}{12}n(n^2-1)\delta_{n-2,0}\right]  W_{-1} +	\\
	& \left. + 3 a_2\left((n+1)W_{n-1} +\frac{c_2}{12}n(n^2-1)\delta_{n-1,0}\right)  W_{-1}^2 \right\}|h\rangle\; ,
	\end{split}
	\ee
	and the expression
	\bse
	\be 
	L_n |\nu\rangle = 0  \quad \forall\; n\in \mathbb{N}^\ast
	\ee
	is, again, identically satisfied for $ h^{(2)}=c_2= 0 $ or if the following conditions are satisfied
	\be 
	h^{(2)}_3 = -\frac{c_2}{3}\; , \qquad a_1=\frac{6}{c_2} \quad \mathrm{and} \quad a_2=0\; .
	\ee
	Obviously, these conditions are consistent with (\ref{null-cond}).
	\ese
Using similar techniques we can determine null states at arbitrary levels.

\begin{shaded}\textbf{Remark:} \emph{Compatibility of all states (and single valuedness of $h^{(2)}$) imposes that $h^{(2)}=c_2=0$. In other words, if one sets the semi-null states to zero, the central charge $c_2$ and the weights $h^{(2)}$  associated to the BMS${}_3$ symmetry are both zero. It is easy to see that the quotient $\mathcal{V}_{h, h^{(2)}}/\mathcal{V}_0$ is, equivalently, defined as a space where the generators $W_m$ act trivially, that is}
\be 
W_n |v\rangle\mapsto 0\; \; \;  \forall n\in \mathbb{Z}\; .
\ee
\emph{Therefore, $\mathcal{V}_{h, h^{(2)}}/\mathcal{V}_0\simeq \mathcal{V}_{h, 0} \simeq \mathcal{V}_{h}$, that is, the final module is the Verma module solely generated by $\{L_{-n}, \ n\geq 1\}$.}
\end{shaded}

In summary, we have seen that for $24 h^{(2)}=-c_2(n^2-1)$, $n\in \mathbb{Z}^\ast$, the Verma module is reducible and that if we want to avoid seminull states, we need to take $c_2=0$. In that case, the resulting module is isomorphic to the states of a Virasoro algebra \cite{Bagchi:2009pe, 2015Jiang}. 

However, we have two strong hypotheses  in our construction\footnote{There is another important hypothesis: the factorization of the CFT into holomorphic and antiholomorphic sectors.}: unitarity and the absence of zero-norm states in the CFT spectrum. Therefore, in order to avoid trivial central charges and W-weights we need to take seriously the existence of seminull states and leave aside the positivity of the Hilbert space.

Additionally, it is obvious that given the level-1 states $|\chi_1 \rangle= W_ {-1} |h \rangle $ and $|\chi_2 \rangle=L _{-1} |h \rangle $ we have the following Jordan structure
\be 
\label{jordan}
W_0 |\chi_1 \rangle=h^{(2)} |\chi_1 \rangle\; , \quad  W_0 |\chi_2 \rangle=h^{(2)} |\chi_2 \rangle + |\chi_1\rangle\; ,
\ee 
where $|\chi_1\rangle$ is evidently a seminull state for generic values of $h^{(2)}$. It is easy to see that this structure also appears for higher-level states; therefore one might try to consider each pair of states, e.g., $(|\chi_1\rangle, |\chi_2 \rangle)$, as ``conjugate'' vectors. We return to all these points in section \ref{nonunit} and propose some solutions to the problems mentioned above.

\section{Correlation functions}

We would like to study the constraints imposed by the BMS${}_3$ extended symmetry to the correlation functions of a unitary CFT. We will see that the Ward identities for two- and three-point functions trivialize the action of the $W_n$ generators. In other words, the results of this section are equivalent to the following statement:

\begin{shaded}{\bf No-go:}
\emph{It is not possible to define a unitary field theory with extended symmetry given by $W(2,2)$ and nontrivial actions of the generators $W_n$.}
\end{shaded}

Let us rewrite the OPE between the algebra generators and primary fields as
\bse
\be 
W^{(\ell)}(z) V_v(\zeta) = \frac{V(W^{(\ell)}_0|v\rangle; \zeta)}{(z-\zeta)^{\ell}} + \frac{V(W^{(\ell)}_{-1}|v\rangle; \zeta)}{(z-\zeta)^{-1+\ell}}+ \cdots + \frac{V(W^{(\ell)}_{1-\ell}|v\rangle; \zeta)}{(z-\zeta)}\; ,
\ee
where 
\be 
V(W^{(\ell)}_{-n}|v\rangle; \zeta)\equiv W_{-n}^{(\ell)} V_v(\zeta)=\frac{1}{2\pi i}\oint_w\dd z\frac{1}{(z-\zeta)^{n+1-\ell}} W^{(\ell)}(z) V_v(\zeta)\; ,
\ee
\ese
and $W^{(1)}(z)\equiv T(z)$ and $W^{(2)}(z)\equiv W(z)$. It is easy to see that the Ward identities are
\bse
\begin{align}
\label{t-ward}
\langle T(z) V_1(z_1)\cdots V_m (z_m)\rangle & = \sum_{k=1}^n \left( \frac{h_k}{(z-z_k)^2} + \frac{\partial_k}{z-z_k} \right) \langle V_1(z_1)\cdots V_m (z_m)\rangle\\
\label{w-ward}
\langle W(z) V_1(z_1)\cdots V_m (z_m)\rangle & = \sum_{k=1}^n \left( \frac{h_k^{(2)}}{(z-z_k)^2} + \frac{W_{-1}}{z-z_k} \right) \langle V_1(z_1)\cdots V_m (z_m)\rangle\; .
\end{align}
\ese
Regularity imposes the conditions
\be 
\lim_{z\to \infty} T(z)=\frac{1}{z^4}\quad \mathrm{and} \quad \lim_{z\to \infty} W^{(2)}(z)=\frac{1}{z^4}\; .
\ee
Using the Ward identities, we can write the correlators for descendants as 
\bse
\begin{align}
\langle L_{-n} V(w) V_1(z_1) \cdots V_m(z_m) \rangle & = \widehat{L}_{-n} \langle V(w) V_1(z_1) \cdots V_m(z_m) \rangle \label{w-eq-01}\\
\langle W_{-n} V(w) V_1(z_1) \cdots V_m(z_m) \rangle & = \widehat{W}_{-n} \langle V(w) V_1(z_1) \cdots V_m(z_m) \rangle  \label{w-eq-02}\; ,
\end{align}
where \footnote{We use the notation 
\begin{displaymath}
{}^kW_{-1}  \langle V_1(z_1)\cdots V_k(z_k)\cdots V_n(z_n)\rangle \equiv  \langle V_1(z_1)\cdots W_{-1}V_k(z_k)\cdots V_n(z_n)\rangle \; .
\end{displaymath}}
\begin{align}
\widehat{L}_{-n} & = (-1)^{n-1}\sum_{k=1}^n \left( \frac{(1-n)h_k}{(w-z_k)^n} - \frac{1}{(w-z_k)^{n-1}} \partial_k \right)\label{op-01}\\
\widehat{W}_{-n} & = (-1)^{n-1}\sum_{k=1}^n \left( \frac{(1-n)h^{(2)}_k}{(w-z_k)^n} - \frac{1}{(w-z_k)^{n-1}} {}^k W_{-1} \right) \label{op-02}\; .
\end{align}
\ese
The modes associated to $W(z)$ are not (at the quantum level) differential operators. Therefore, given a null vector $|\chi_{h^{(2)},c_2}\rangle$ associated to highest-weight state $| h, h^{(2)}\rangle$, the constraints that this state imposes to the correlation functions are not, contrary to the usual Virasoro null states, differential equations. For example, using the results of the previous section, the null states at level 2 and 3 give, respectively, the two conditions
\be
\begin{split}
\langle \chi(z) \phi_1(w_1)\cdots\rangle &= \left( W_{-2} +\frac{6}{c_2} W_{-1}^2\right) \langle V_{h_1}(z) \phi_1(w_1)\cdots \rangle = 0 \\
\langle \mu(z) \phi_1(w_1)\cdots \rangle & = \left( W_{-3} +\frac{6}{c_2} W_{-2}W_{-1}\right) \langle V_{h_1}(z) \phi_1(w_1)\cdots \rangle = 0 \; ,
\end{split}
\ee
which in present state seem to be completely algebraic.

Let us assume the existence of an $SL(2, \mathbb{C})\cup \{ W_{-1}, W_0, W_{+1} \}$ invariant vacuum $|0\rangle$. Using $V(w)=V(|0\rangle;w)=\mathbb{1}$ and taking $w=0$ we have the equations
\bse
\begin{align}
& \sum_{k=1}^n \partial_k G_m  = 0\; , \quad
\sum_{k=1}^n\left(h_k + z_k \partial_k \right) G_m  = 0\; , \quad
\sum_{k=1}^n\left(2 z_kh_k + z_k^2 \partial_k \right) G_m  = 0 \label{op-01a}\\
& \sum_{k=1}^n {}^kW_{-1} G_m  = 0\; , \quad
\sum_{k=1}^n\left(h_k^{(2)} + {}^kW_{-1} z_k  \right) G_m  = 0\; , \quad
\sum_{k=1}^n\left(2z_kh_k^{(2)} + {}^kW_{-1} z_k^2 \right) G_m  = 0\; , \label{op-02a}
\end{align}
\ese
where $G_m = \langle V_1(z_1) \cdots V_m(z_m)\rangle$.

The $T$-Ward identities impose the usual conditions for the one-, two- and three-point functions, and the $W$-Ward identities impose further restrictions that we shall verify now. The one-point functions are zero, so let us start with the two-point functions.

\paragraph{Two-point functions:} Equations (\ref{op-02a}) can be organized as the  linear system
\be
\label{ward-matrix}
 R\cdot \vec{V} = 
\begin{pmatrix}
0 & 1 & 1\\
h^{(2)}_1 + h^{(2)}_2 & z_1 & z_2 \\
2(h^{(2)}_1 z_1 + h^{(2)}_2 z_2) & z_1^2 & z_2^2\\
\end{pmatrix}
\begin{pmatrix}
\langle V_1(z_1) V_2(z_2)\rangle\\
\langle (W_{-1} V_1(z_1)) V_2(z_2)\rangle\\
\langle V_1(z_1) (W_{-1} V_2(z_2))\rangle\\
\end{pmatrix} = 0\; ,
\ee
and nontrivial solutions exist iff $\det R=-\left(h^{(2)}_1 - h^{(2)}_2\right)(z_1-z_2)^2=0$. Therefore, $h^{(2)}_1 = h^{(2)}$, and we conclude that the two-point functions are
\be 
\langle V_1(z_1) V_2(z_2) \rangle= \frac{c_{12} \delta_{h_1 h_2} \delta_{h^{(2)}_1 h^{(2)}_2}}{z_{12}^{2h_1}}\; .
\ee
This condition also appears in \cite{Fateev:2007ab, Fateev:2008bm} for the ${\cal W}_3$ algebra, except that in the present case the $W_0$ eigenvalues $h_k^{(2)}$ must be equal.

Given that $h^{(2)}_1=h_2^{(2)}\equiv h^{(2)}$ must be satisfied in order to get a nontrivial correlation function, the first condition in the Ward identity (\ref{ward-matrix}) is
\bse
\be 
\langle W_{-1} V_1(z_1) V_2(z_2) \rangle = - \langle V_1(z_1) W_{-1} V_2(z_2) \rangle\; ,
\ee
and the remaining equations form the system
\be 
\begin{split}
& 2 h^{(2)}\langle V_1(z_1) V_2(z_2) \rangle + z_1 \langle W_{-1} V_1(z_1) V_2(z_2) \rangle + z_2 \langle W_{-1} V_1(z_1) V_2(z_2) \rangle=0\\
& 2 h^{(2)}(z_1 + z_2)\langle V_1(z_1) V_2(z_2) \rangle + z_1^2 \langle W_{-1} V_1(z_1) V_2(z_2) \rangle + z_2^2 \langle W_{-1} V_1(z_1) V_2(z_2) \rangle=0\; .
\end{split}
\ee
Putting all these facts together, we find
\be 
\langle V_1(z_1) W_{-1}  V_2(z_2) \rangle = \frac{h^{(2)} c_{12} \delta_{h_1 h_2} \delta_{h_1^{(2)} h_2^{(2)}} }{z_{12}^{2h_1 + 1}}\; .
\ee
\ese

\paragraph{Three-point functions:} Let us turn our attention to the three-point functions. From (\ref{op-02a}), one gets the linear system
\bse
\begin{align}
& \sum_{k=1}^{3} {}^kW_{-1}  \langle V_1(z_1) V_2(z_2) V_3(z_3)\rangle = 0\label{eq01}\\
& \sum_{k=1}^{3}\left(h^{(2)}_k +  {}^kW_{-1} z_k \right)\langle V_1(z_1) V_2(z_2) V_3(z_3)\rangle = 0\\
& \sum_{k=1}^{3}\left(2 h^{(2)}_k z_k + {}^kW_{-1} z_k^2  \right)\langle V_1(z_1) V_2(z_2) V_3(z_3)\rangle = 0\; . \label{eq03}
\end{align}
\ese
It is an underdetermined system: three equations and four unknowns, that is, $\langle V_1 V_2 V_3\rangle$, $\langle W_{-1}V_1 V_2 V_3\rangle$, $\langle V_1 W_{-1}V_2 V_3\rangle$ and $\langle V_1 V_2 W_{-1} V_3\rangle$. 
\begin{itemize}
	\item[{\it i})] The first condition obviously gives
	\be 
	\langle V_1(z_1) V_2(z_2) W_{-1} V_3(z_3)\rangle = - \langle W_{-1} V_1(z_1) V_2(z_2) V_3(z_3)\rangle - \langle V_1(z_1) W_{-1} V_2(z_2) V_3(z_3)\rangle\; .
	\ee
	\item[{\it ii})] Inserting the last expression into the second and third equations of the linear system gives
	\bse
	\be 
	\begin{split}
	\left[(z_1 - z_3) ({}^1W_{-1})  + (z_2 - z_3) ({}^2W_{-1}) \right] & \langle V_1(z_1) V_2(z_2) V_3(z_3)\rangle =\\
	& = - (h_1^{(2)} + h_2^{(2)} + h_3^{(2)})\langle V_1(z_1) V_2(z_2) V_3(z_3)\rangle
	\end{split}
	\ee
	and 
	\be 
	\begin{split}
	\left[(z_1^2 - z_3^2) ({}^1W_{-1})  + (z_2^2 - z_3^2) ({}^2W_{-1}) \right] & \langle V_1(z_1) V_2(z_2) V_3(z_3)\rangle =\\
	& = - 2 (h_1^{(2)}z_1 + h_2^{(2)}z_2 + h_3^{(2)}z_3 )\langle V_1(z_1) V_2(z_2) V_3(z_3)\rangle \; ,
	\end{split}
	\ee
	\ese
	respectively.
	\item[{\it iii})] From these last conditions, we find
	\be 
	\begin{split}
	\langle(W_{-1} V_1(z_1)) V_2(z_2) & V_3(z_3)\rangle \equiv \Delta(z_k ;h_k^{(2)}) \langle(V_1(z_1)) V_2(z_2) V_3(z_3)\rangle \\
	& = \frac{C_{123}\left(  2( h_{1}^{(2)} z_1+ h_{2}^{(2)} z_2 +h_{3}^{(2)} z_3) + (z_2 + z_3)(h_{1}^{(2)}+h_{2}^{(2)} + h_{3}^{(2)} )\right)}{z_{12}^{h_{12} + 1} z_{13}^{h_{13}+ 1} z_{23}^{h_{23}}} \; ,
	\end{split}
	\ee
	where we have used the textbook result for the three-point function \cite{Ketov:1995yd, DiFrancesco:1997nk, Blumenhagen:2009zz}.
\end{itemize}

Using these three expressions, we can write the equations (\ref{op-02a}) as the homogeneous system
\be
S\cdot \vec{U} = 
\begin{pmatrix}
\Delta & 1 & 1\\
h^{(2)}_1 + h^{(2)}_2  + h^{(2)}_3 + z_1 \Delta & z_2 & z_3 \\
2(h^{(2)}_1z_1 + h^{(2)}_2 z_2  + h^{(2)}_3 z_3) + z_1^2 \Delta & z_2^2 & z_3^2\\
\end{pmatrix}
\begin{pmatrix}
\langle V_1(z_1) V_2(z_2)V_3(z_3)\rangle\\
\langle V_1(z_1)  (W_{-1} V_2(z_2)) V_3(z_3)\rangle\\
\langle V_1(z_1)V_2(z_2)  (W_{-1} V_3(z_3))\rangle\\
\end{pmatrix} = 0\; ,
\ee
and again, taking the determinant of $S$ and setting it to zero, we have
\be 
-4 (z_2-z_3)(h^{(2)}_1z_1 + h^{(2)}_2 z_2  + h^{(2)}_3 z_3) = 0\; ,
\ee
which must be satisfied for any value of $z_1, z_2$, and $z_3$; then,
\be 
h_1^{(2)}= h_2^{(2)}= h_3^{(2)}=0\; .
\ee
Therefore, the action of the $W_n$ modes in a unitary CFT algebra is trivialized.

\section{Nonunitary CFTs}\label{nonunit}

As we have seen in the previous sections, if assume that the BMS${}_3$ invariant CFTs are unitary, we find very unpleasant results. First of all, we cannot generate the
\(W(z)\) operators with nontrivial central charges \(c_2\) following a modification of the Sugawara construction based on semisimple Lie algebras. Additionally, the null states and the correlation function analysis indicate that the action of this symmetry algebra is trivial.

One important aspect of our study is the presence of zero-norm vectors, called \emph{seminull}, which are not orthogonal to the whole Verma module. We know that in a unitary CFT, null states generate their own Verma module, and that the set of ``physical'' vectors is drastically reduced. In other words, we take the quotient of the whole Verma module by the space of null states, that is, $\mathcal{V}_{(h,c)}/\mathcal{V}_0$.

Unitarity is, evidently, an important property in fundamental theories, but we must be careful with its role, and not overestimate its importance in general CFTs. First of all, nonunitary theories have their own place in the realm of condensed matter systems \cite{Mussardo:2010mgq}. Additionally, unitarity of the vector space of physical states is much more important than the unitarity of the theory itself. For example, the gauge fixed plus ghost action is generally nonunitary (evidently, this action is not obtained from first principles); on the other hand, the BRST cohomology must, evidently, be positive definite.

It is not clear at the present stage whether the presence of the seminull states necessarily gives a nonunitary CFT, but let us assume that it is the case. Therefore, we would like to discuss the resulting aspects of nonunitarity in our analysis; that is, we would like to see the results the condition $h^{(2)}\neq 0$. Consider again the states $|\lambda_1^{(N)} \rangle=W_{-N}|h\rangle$ and $|\lambda_2^{(N)}\rangle =L_{-N}|h\rangle$. The Jordan structure (\ref{jordan}) is generalized to 
\be 
\label{jordan-N}
\begin{split}
&  L_0 |\lambda_1^{(N)} \rangle = (h+N) |\lambda_1^{(N)} \rangle\; , \quad L_0 |\lambda_2^{(N)} \rangle = (h+N) |\lambda_2^{(N)} \rangle\\
&  W_0 |\lambda_1^{(N)} \rangle = h^{(2)}|\lambda_1^{(N)} \rangle\; , \quad W_0 |\lambda_2^{(N)} \rangle = h^{(2)} |\lambda_2^{(N)} \rangle+ N |\lambda_1^{(N)} \rangle\; .
\end{split}
\ee
Given an infinite set of nilpotent variables $\{\theta_n\ |\ \theta_n^2=0\; , \forall \ n\in \mathbb{Z}\}$, we define the state
\bse
\be 
\begin{split}
|\Phi_N\rangle := & |h_N; h^{(2)}+\theta_N\rangle\\
= & |h_N, h^{(2)}\rangle_0 +\theta_N |h_N, h^{(2)}\rangle_1\\
\equiv & |\lambda_1^{(N)} \rangle +\theta_N |\lambda_2^{(N)} \rangle\; ,
\end{split}
\ee
where in the second line we have written the Taylor expansion in $\theta_N$. Let us say that $|\lambda_2^{(N)} \rangle$ is the Virasoro state and $|\lambda_1^{(N)} \rangle$ is its seminull partner. Hence, we rewrite the Jordan structure (\ref{jordan-N}) as
\be 
L_0 |\Phi_N\rangle = (h+N) |\Phi_N\rangle\; , \quad W_0 |\Phi_N\rangle = (h^{(2)}+ N \theta_N) |\Phi_N\rangle\; .
\ee

Evidently, we still have states of the form $L_{-\vec{n}}W_{-\vec{m}}|h\rangle$, where $N=|\vec{n}|+|\vec{m}|$, but using the commutation relations, we can write all these cross terms as descendants of $|\lambda_1^{(N')} \rangle$ and $|\lambda_2^{(N')} \rangle$ with $N'<N$ in the following way: at each level $N$, let us define the operators $\mathbb{W}^{(N)}_n$ as
\be 
\mathbb{W}_n^{(N)}:= W_n + N \theta_N L_n\; ,
\ee
where the role of the index $n=1, \cdots,N $ will be explained soon. Observe that the state $|\Phi_N\rangle$ is written as
\be 
\label{states}
|\Phi_N\rangle: = \mathbb{W}_{-N}^{(N)} |h\rangle\; .
\ee
\ese

Let us consider the first states. It easy to see that $|\Phi_0\rangle= |h\rangle$. At level 1, we have $n=1$; therefore,
\be 
|\Phi_1\rangle= \mathbb{W}^{(1)}_{-1} |\Phi_0\rangle = |\lambda_1\rangle+ \theta_1 |\lambda_2\rangle\; ,
\ee
where $|\lambda_1\rangle= W_{-1}|\Phi_0\rangle$ and $|\lambda_2\rangle =L_{-1}|\Phi_0\rangle$ have conformal weights $h_1=h+1$. Therefore
\be 
\begin{split}
L_0 |\lambda_1\rangle & =  h_1 |\lambda_1\rangle\;  \qquad L_0 |\lambda_2\rangle =  h_1 |\lambda_2\rangle\\
W_0 |\lambda_1\rangle & =  h^{(2)} |\lambda_1\rangle \qquad W_0 |\lambda_2\rangle =  h^{(2)} |\lambda_2\rangle + |\lambda_1\rangle\; .
\end{split}
\ee
We see that the operator $W_0$ cannot be diagonalized and we can say that $|\lambda_2\rangle$ is the Jordan partner of $|\lambda_1\rangle$. Finally, we already know that $L_1 |\lambda_1\rangle=2h^{(2)}|h\rangle$.

At level 2, we have the operators $\mathbb{W}^{(2)}_{-1}$ and $\mathbb{W}^{(2)}_{-2}$. From (\ref{states}), one has
\be 
|\Phi_2\rangle= \mathbb{W}^{(2)}_{-2} |\Phi_0\rangle = |\lambda^{(2)}_1\rangle+ 2 \theta_2 |\lambda^{(2)}_2\rangle\; ,
\ee
where $|\lambda^{(2)}_1\rangle= W_{-2}|\Phi_0\rangle$ and $|\lambda^{(2)}_2\rangle =L_{-2}|\Phi_0\rangle$. Evidently, there are other states at level 2 in the Verma module. The remaining fields can be built by applying the operator $\mathbb{W}^{(2)}_{-1}$ in the state $|\Phi_1\rangle=\mathbb{W}^{(1)}_{-1} |\Phi_0\rangle$, that is,
\be 
\mathbb{W}^{(2)}_{-1}|\Phi_1\rangle = \left( W_{-1}^2 + \theta_1 W_{-1}L_{-1} +2 \theta_2 L_{-1}W_{-1} + 2 \theta_2 \theta_1 L_{-1}^2\right)|\Phi_0\rangle\; .
\ee
Therefore, all level-2 Verma module is organized in the multiplet $\{|\Phi_2\rangle, \mathbb{W}^{(2)}_{-1}|\Phi_1\rangle\}$. Similarly, the level-3 can be organized in $\{|\Phi_3\rangle, W_{-1}^{(3)}|\Phi_2\rangle, W_{-1}^{(3)} W_{-1}^{(2)} |\Phi_1\rangle\}$. All in all, we see that the Virasoro states and their seminull partners can be seen as the building blocks of this construction and that we can organize all Verma module in terms of the set $\{W_n^{(N)}, |\lambda_1^{(N)}\rangle, |\lambda_2^{(N)}\rangle\}$. Evidently, we have some redundancies to be removed, for example $L_{-n}W_{-m}|\Phi_0\rangle$ and $W_{-m}L_{-n}|\Phi_0\rangle$ for $N=n+m$ that are equivalent modulo the state $W_{-N}|\Phi_0\rangle$.

Finally, using the operator-state map we define
\bse
\be 
\Phi_N(z)=\varphi_N(z)+ \theta_N \psi_N(z)\; \Rightarrow\; |\Phi_N\rangle=\lim_{z\to 0} \Phi_N(z) |h\rangle\; ,
\ee
so that 
\be 
|\lambda^{(N)}_1\rangle = \lim_{z\to 0} \varphi_N(z)|h\rangle\; , \quad |\lambda^{(N)}_2\rangle = \lim_{z\to 0} \psi_N(z)|h\rangle\; .
\ee
\ese

The similarity between this theory and the structures of the Logarithm CFTs (LogCFTs) \cite{Gaberdiel:2001tr, Flohr:2001zs, Creutzig:2013hma} has been pointed out in \cite{Bagchi:2009pe, Bagchi:2012yk, 2016Adamovic, Adamovic2017, Henkel:2013ecs}, but the relation between them is not straightforward, and as the results of this section show, the Jordan structure in the present case is associated to the seminull descendant fields. The BMS${}_3$ invariant CFTs are much more similar to the theories considered in \cite{Golkar:2014mwa, Brust:2016gjy}. In that case, the authors argued that, contrary to the unitarity CFTs, the existence of zero-norm states does not mean the reduction of the Verma module and that its degeneracy, that is, the presence of seminull states should be compensated by the some \emph{extension} (or \emph{alien}) states $|\tilde{\chi}\rangle$ such that
\be 
\langle \tilde{\chi} | \chi\rangle \neq 0\; ,\qquad \langle \tilde{\chi} | \tilde{\chi}\rangle \neq 0\; ,
\ee
where $|\tilde{\chi}\rangle$ is a ``partner'' of the zero-norm state $|\chi\rangle$. 

Then, for each seminull state $|\chi\rangle$, we have an extended vector $|\tilde{\chi}\rangle$ that generates its own module and needs to be appended to the initial space of states. But this is exactly what we have found in our computations and the only noticeable difference between the present case and \cite{Golkar:2014mwa, Brust:2016gjy}, is that our zero-norm states are associated with descendant fields, while their construction associates to the seminull states, extension operators that are neither descendant nor primaries. Despite this small difference, the seminull partners (extension operators) are natural features of our constructions, and not mere \emph{ad hoc} objects.

In particular, it is obvious that the stress-energy tensor $T(z)$ and the BMS${}_3$ generator $W(z)$ are descendants. Given the states $|\mathsf{T} \rangle = L_{-2} |0\rangle$ and $|\mathsf{W} \rangle = L_{-2} |0\rangle$, their vertex operators are
\be 
T(z):=V(|\mathsf{T} \rangle;z)\quad \mathrm{and} \quad W(z):=V(|\mathsf{W} \rangle;z)\; .
\ee
Therefore, we have the pair
\be 
{\cal T}(z):= W(z) + \theta T(z)\; ,
\ee
and we see that the fields $(T(z), W(z))$ belong to the same ``multiplet''.

We conclude that the symmetry transformations generated by the BMS${}_3$ algebra must be considered simultaneously. As we have seen in the previous section, the W-ward identities impose severe constraints to our theory, namely $h^{(2)}=0$, and now we understand this fact as a consequence of our naive attempt to consider the symmetries associated to $T(z)$ and $W(z)$ as independent transformations. Consequently, we need to modify the Ward identities. 

Let us start modifying the OPE $W(z) V_h(\zeta)$, that is
\begin{align}
W(z) V_h(\zeta) & = \frac{h^{(2)} V_{h}(\zeta) }{(z-\zeta)^2}  +  \frac{(W_{-1}V_{h})(\zeta) +(\partial V_h)(\zeta) }{(z-\zeta)} \; ;
\end{align}
hence, the operator (\ref{op-02}) becomes
\be 
\widehat{W}_{-n} = (-1)^{n-1}\sum_{k=1}^n \left( \frac{(1-n)h^{(2)}_k}{(w-z_k)^n} - \frac{1}{(w-z_k)^{n-1}} ({}^k W_{-1}+\partial_k) \right)\; .
\label{op-02m}
\ee
Again, this is similar to the LogCFT, where we modify the Ward identity by a nilpotent operator \cite{Flohr:2001zs, Creutzig:2013hma}. 

Let us reconsider the expressions (\ref{w-eq-02}) and (\ref{op-02a}). Using (\ref{op-02m}); imposing the invariance of the vacuum under $\{L_n, W_n| n=-1,0,+1\}$, that is $\widehat{W}_{-n} G_m=0$, and finally using the equations (\ref{op-01a}), we have 
\bse
\begin{align}
& \sum_{k=1}^n {}^kW_{-1} G_m  = 0\\
& \sum_{k=1}^n\left(h_k^{(2)}-h_k + {}^kW_{-1} z_k  \right) G_m  = 0\\
& \sum_{k=1}^n\left(2z_k(h_k^{(2)}-h_k) + {}^kW_{-1} z_k^2 \right) G_m  = 0\; .
\end{align}
\ese
With this new identity, we can repeat the computations of the previous section for the two-point functions to see that they are not modified. For the three-point functions we find the condition $h_k^{(2)}=h_k$.

The nice aspect now is that we can compute four-point functions using the usual technology we have in any CFT with Virasoro symmetry. The four-point function is written in terms of conformal blocks
\be 
G_4(x)= \sum_p C_{12}^p C_{34}^p \mathcal{F}(p|x)\; , \quad x=\frac{z_{12} z_{34}}{z_{13} z_{24}}\; ,
\ee
and the crossing symmetry can be imposed via the bootstrap approach \cite{DiFrancesco:1997nk, Ketov:1995yd}. Evidently, the conditions imposed by the W-Ward identities are ``essentially'' algebraic, given that we do not have a clear understanding of the nature of the operators $W_{n}$ in a general ${\cal W}$ algebra.

We have dealt with the existence of the seminull states by a simple modification of the W-Ward identity. But evidently, we need to get rid of the proper null states, since these fields must be absent in the final Verma module. In practice, we are very familiar with these null states, and we know from many standard examples, see \cite{Ketov:1995yd, DiFrancesco:1997nk, Blumenhagen:2009zz}, that these singular vectors give further conditions to the correlation functions. 

For example, using the state (\ref{ba-null}) and supposing that it is indeed null, its associated vertex operator $\chi(z)$ can be inserted in a generic correlation function. Then,
\be 
\left(  a_1 \widehat{W}_{-2} + a_2 \widehat{L}_{-2} + a_3 \widehat{L}_{-1} \widehat{W}_{-1} + a_4 \widehat{W}_{-1}^2 + a_5 \widehat{L}_{-1}^2 \right)\langle V(w) V_1(z_1)\cdots V_m(z_m) \rangle  = 0\; .
\ee
Consider now that the W weight is given by $24 h^{(2)} = -c_2 (n^2-1)$ for some $n\in \mathbb{Z}$; then null vector is unique \cite{2015Jiang}. For a particular example, we assume that $n=2$; then, the condition 
\be 
\left(\widehat{W}_{-2} + \frac{6}{c_2} \widehat{W}_{-1}^2  \right)\langle V(w) V_1(z_1)\cdots V_m(z_m) \rangle  = 0
\ee
is defined by the state (\ref{null-st2}).

\section{Conclusions and further directions}

In this paper, we explored the equivalence BMS${}_3\simeq W(2,2)$ to remark some aspects of meromorphic CFTs with extended chiral  symmetries given by the BMS${}_3$ algebra. Initially, we tried to study the structure of these generic theories using the results and insights of unitary theories, for example, the Sugawara construction. The drawback of this approach is that it the algebraic aspects of the nonsemisimple Lie algebra are, at the present stage, very elusive.

To study these theories, we have seen that there are certain zero-norm vectors that are not necessarily annihilated by the modes $\{L_n, W_m\}$, with $n,m>0$. The removal of these states provided us with a relation between the $W$ weights $h^{(2)}$ and the central charge $c_2$. In the particular case of vanishing central charge, $c_2=0$, it is well known that the Verma modules for these theories are equivalent to the Verma modules of ordinary CFTs \cite{2015Jiang}. 

We turned our attention to the analysis of Ward identities, which are, in a very precise way, the quantum manifestation of the classical symmetries of the system, and they are the cornerstone of any consistent QFT with classical local or global symmetries. Furthermore, the Ward identities for the $W$ symmetry imply that the weights of all states must be zero; therefore, we cannot have a unitary field theory with a nontrivial $W$ symmetry action.

Next, we considered nonunitary CFTs and we assumed that seminull states are important in our setting. It is not known whether the presence of seminull states necessarily gives a nonunitary CFT, and it would be interesting to check this condition carefully. In any case, the presence of these states gives a rich structure to the theory, and we organized the Verma module through a nilpotent operator that mixes the BMS${}_3$ generators. Moreover, we have shown that in the nonunitary setting the W-Ward identities do not trivialize the action of the  symmetry operators.

There are many directions for future research. One can try to perform an mathematical analysis of the conditions we considered in section \ref{nonunit}, in particular, one can try to understand the similarities between the BMS${}_3$ theories and LogCFTs. Additionally, since we do not have an interpretation for the operators $W_{n}$, there is no good reason to believe that these operators respect the holomorphic and antiholomorphic factorization of the theory (for example, these operators could act as a multiplication by an antiholomorphic function); therefore one could try to relax this condition.

One can try to change the OPE $W(z)W(\zeta)$ and see if one can find a consistent CFT with extended algebra, or add-higher spins operators in this construction and see how the Verma module is modified. Remember that in the usual $\mathcal{W}_{\ell}$ algebra approach, one considers just one additional primary operator $W^{(\ell)}(z)$ for each integer $\ell\geq 3$. Then, the study we considered in this paper is part of a bigger problem: the study of CFTs with extended symmetries associated with several quasiprimary operators of the same conformal dimension.

It is known that the $\mathcal{W}_\ell$ algebras are the symmetry of the A${}_{\ell-1}$ Toda systems, and in particular, the Virasoro algebra $\mathcal{W}_2\equiv \texttt{Vir}$ is the symmetry algebra of the simplest Toda system, the Liouville CFT, see \cite{Teschner:2001rv, Fateev:2007ab, Fateev:2008bm} and references therein. In \cite{Barnich:2012rz}, the relation between some ``flat space limit'' of the Liouville theory and BMS${}_3$ symmetry has been analyzed; therefore, it would be interesting to see if we can connect the present construction to some deformation of the Liouville CFT. Even more speculatively, it would be interesting to see if conformal blocks of these elusive theories are connected to the partition functions of four-dimensional theories, \emph{\`a  la} AGT correspondence \cite{Alday:2009aq, Wyllard:2009hg}. We hope to address some of these points in a future work.

\section*{Acknowledgments}

I would like to thank Shahin Sheikh-Jabbari, Eoin \'O Colg\'ain, Thiago Fleury, and Carlos Cardona for discussions, correspondence, and critical reading of the manuscript. I would also like to thank the anonymous referees for valuable comments, criticism, and suggestions. This work was supported by the Korea Ministry of Education, Science and  Technology, Gyeongsangbuk-Do and Pohang City Independent Junior Research Groups at the Asia Pacic Center for Theoretical Physics.

\addcontentsline{toc}{section}{References}

{\small
\bibliographystyle{utphys}
\bibliography{library.bib}
} 
 
\end{document}